# Thermodynamic Description of Beta Amyloid Formation

J. C. Phillips[2]

Dept. of Physics and Astronomy, Rutgers University, Piscataway, N. J., 08854

## Abstract

Protein function depends on both protein structure and amino acid (aa) sequence. Here we show that modular features of both structure and function can be quantified from the aa sequences alone for the small (40,42 aa) plaque-forming amyloid beta fragments. Some edge and center features of the fragments are predicted. Contrasting results from the second order hydropathicity scale based on evolutionary optimization (self-organized criticality) and the first order scale based on complete protein (water-air) unfolding show that fragmentation has mixed first- and second-order character.

The precursor amyloid protein A4 of Alzheimer's disease (AD) (details available on Uniprot as P05067) has 770 aa and resembles a cell surface receptor. Two small amyloid fragments of A4 – Aβ 40, 672-711, and Aβ 42, 672-713 - form plaques. Monomer fragments aggregate to form successive oligomeric globules, proto-fibers, or fibers, with the latter being most toxic. An NMR structural model [1] shows that the cores of these fragments are dominated by a sandwich structure with halves, β1 = 18-26 and β2 = 31-40 or 31-42, connected by a β turn stabilized by an E22,D23-K28 salt bridge. A hydrophilic loop 1-17 connects successive monomers (Fig.1, from [1]).

Presently evidence suggests that oligomer intermediates, which lead to fibril formation, may be the primary culprits leading to neuronal dysfunction and cell death [2]. Numerous studies have [3-10], suggested that Aβ 42 is more toxic than Aβ 40. The molecular origin of the biological differences between Aβ 40 and 42 is still under study [2,11].

Here we examine nonlocal hydropathic Aβ 40 and 42 interactions within a modular framework based on the concept that evolution has created self-organized protein structures that are thermodynamically nearly critical [12,13]. Our folding analysis, based on amino acid hydropathic interactions, is simple, economical, and quantitative. Its universality extends well beyond amyloid fragments and plaque disruption of neural networks. For example, oligomer formation is critical to viral activity, and has been the subject of > 3000 articles. The present convoluted (long-range) method has already been applied to both small (140 aa) lysozyme [13] and large (~ 350 aa) GPCR membrane [14,15] and (~500 aa) viral proteins [16], as well as the very large 770 aa amyloid precursor A4 [17]. By including hydropathic interactions, which distinguish Aβ 40 and 42, we are able to extract the central modular features of the $\beta_1 - \beta_2$ sandwich by analyzing only aa sequences.

In earlier papers [13,14,16,17] we discussed the thermodynamic significance of the standard water-air, first-order, enthalpy-based KD hydropathicity scale [18]. The new MZ hydropathicity scale, based on self-organized criticality (SOC), was derived from the differential geometry of > 5000 protein segments [12]. Its derivation strongly suggests second-order character, suitable for describing modular conformational activity, which should dominate most protein functionality. Earlier results from correlations between evolutionary trends in functionalities and aa sequences generally showed better agreement using the MZ scale compared to the KD scale.

Earlier work comparing the two ψ(aa,1) scales often examined the variance (or roughness) ℜ of modular sequences ψ(aa,W), where ψ(aa,W) is ψ(aa,1) averaged over a sliding window of width W. The intuitive interpretation of W is that it represents the length of a modular element that dominates the protein functionality. The nature of this modular element is especially clear for the amyloid precursor protein A4, whose critical step is fragmentation into Aβ 40 or Aβ 42, here often renumbered as 1-40 or 1-42, respectively. We were able to identify a thermodynamic spinodal by identifying two closely spaced weak breaks in slope of the KD ℜ(ψ(aa,W)), and two widely spaced larger breaks in slope of the MZ ℜ(ψ(aa,W)). This shows that the MZ scale corresponds to a lower effective temperature than the KD scale. It explains why the MZ scale gave better correlations between evolutionary trends in functionalities and aa sequences than the KD scale for many other proteins.

Given the A4 spinodal, we also saw that the modular width W is analogous to the molar volume $V_m$, and $\Re$ is analogous to the pressure p in gas-liquid equations of state, such as the van der Waals (vdW) equation. Modularity emerges spontaneously in aa distributions of primitive proteins [19]. In modern proteins its effects can be identified using thermodynamic variables W and $\Re$ derived from aa using hydropathic scales.

The amyloid fragments Aβ occur in the 40 aa left half of the compound C-terminal hydrophobic peak [17]. This 80 aa peak is strongly hydrophobic, and rises well above the profile of the rest of A4. It is also about four times as long as a typical 20 aa transmembrane segment of a GPCR protein [14]. It splits in half to form the Aβ fragments, which then self-organize into plaque.

The detailed KD and MZ W = 21 profiles show how this occurs (Fig. 2). As noted in [17], W = 21 is a natural choice for W in the vdW equation of state, but it is especially suited to describing A4 fragmentation between 671 and 672 by the membrane-bound enzyme secretase family [20]. Both scales show that the Aβ fragments are strongly amphiphilic [21], with the left end located at a hydrophilic minimum for W = 21, and the right end at a hydrophobic maximum.

The MZ success achieved with W = 21 in predicting the 672 and 712±1 cleavage sites is unique. Fig. 3 shows the results with other choices of W. These yield extrema near, but not at, the experimental cleavage sites. Profile extrema drift due to numerical noise is natural in such complex proteins, which suggests that the successes achieved with W = 21, and particularly the MZ scale, is thermodynamic in nature and derives from the low effective temperature associated with the MZ scale.

The variance or roughness $\Re(W) = \langle[\psi(aa,W)]^2\rangle - [\langle\psi(aa,W)\rangle]^2$ measures the effective pressure p exerted by the "ball bearing" roughness of the protein's water package as it slides and/or tumbles into successive conformations. A simple macroscopic model for the hydrophobic β sandwich shown in Fig. 1 supposes that the β1 and β2 strands are stabilized both by the β turn and by velcro-like roughness of intra- and inter-sandwich packing under the influence of water pressure. In other words, the larger is the roughness, the greater the sandwich, oligomer, and fibril thermal stabilization.

The velcro analogy is easily tested for the small β amyloid fragments by plotting ℜ(W) for more toxic Aβ 42, normalized by ℜ(W) for less toxic Aβ 40, using either the SOC MZ scale [12], or the water-air unfolding scale KD [18]. As shown in Fig. 4, hydropathic modularity is unambiguous, with peaks at W = $W_{max}$ = 13, and harmonics at W = 27, with both MZ and KD scales. It is striking that the differential MZ scale based on SOC (second-order phase transition) is twice as effective as the complete unfolding (first-order) KD scale in identifying $W_{max}$. This confirms previous results for the MZ superiority found in other larger proteins [13,14,16]. It is easy to see that $W_{max}$ should be 13, as this divides Aβ into three nearly equal parts, the hydrophilic loop, and the two hydrophobic $\beta_1 - \beta_2$ strands.

The differences between the KD complete unfolding scale, and the MZ differential or conformational scale, are neither accidental nor incidental. The associated profiles of ψ(aa,13) are shown in Fig. 5. The MZ scale reflects the differences between the hydrophilic loop (1-17) and the hydropathic sandwich strands 18-42 more accurately, which in turn more accurately resolves the modular structure functionality, including even special hydrophilicity for the hairpin turn, with the MZ scale. Similar plots for W = 9 and W = 17 (not shown), are less informative, as expected from Fig. 2.

In discussing A4 (770 aa) we used spinodal breaks to argue [17] that the MZ scale measures fragmentation at a lower effective temperature. One can still define effective temperatures for the Aβ 40/42 fragments from the values of ℜ(W) for the two scales. This gives effective temperature or ℜ(W) ratios of KD/MZ ranging from 1.67 (W = 13) to 1.93 (W = 21). These ℜ(W) ratios are sensible for two reasons: (1) the effective temperatures of the KD scale are higher than those for the MZ scale, and (2) fragmentation is best described by W = 21, where the ℜ(W) ratio is larger and the effective temperatures are higher, and sandwich reconstruction and aggregation is best described by W = 13 (Figs. 4 and 5), where the ℜ(W) ratio is smaller and the effective temperatures are lower.

There are many Aβ familial mutations, and even in vitro these exhibit a wide range of oligomer and fibrous morphologies [1]. Some of these can be explained by roughness changes, with success similar to that achieved by other means, including atomic force microscopy and molecular dynamics. The A673T mutation that reduces β cleavage [22] broadens the hydrophilic minimum centered on Asp 672 (Fig. 2), which could explain reduced effectiveness of secretase cleavage.

Our analysis of the β fragments, which are only ~ 40 aa long, has confirmed the central conclusion of our earlier analysis of A4 (770 aa). Simple algebraic analysis, using only amino acid sequences, has revealed thermodynamically significant aspects of plaque formation. In the present case, an amphiphilic Velcro model has emerged for the β amyloid fragments that may encompass some other β strand structures as well. We have also found consistent values of effective temperatures, which are frequently used to describe the properties of glasses and deeply supercooled liquids [23,24].

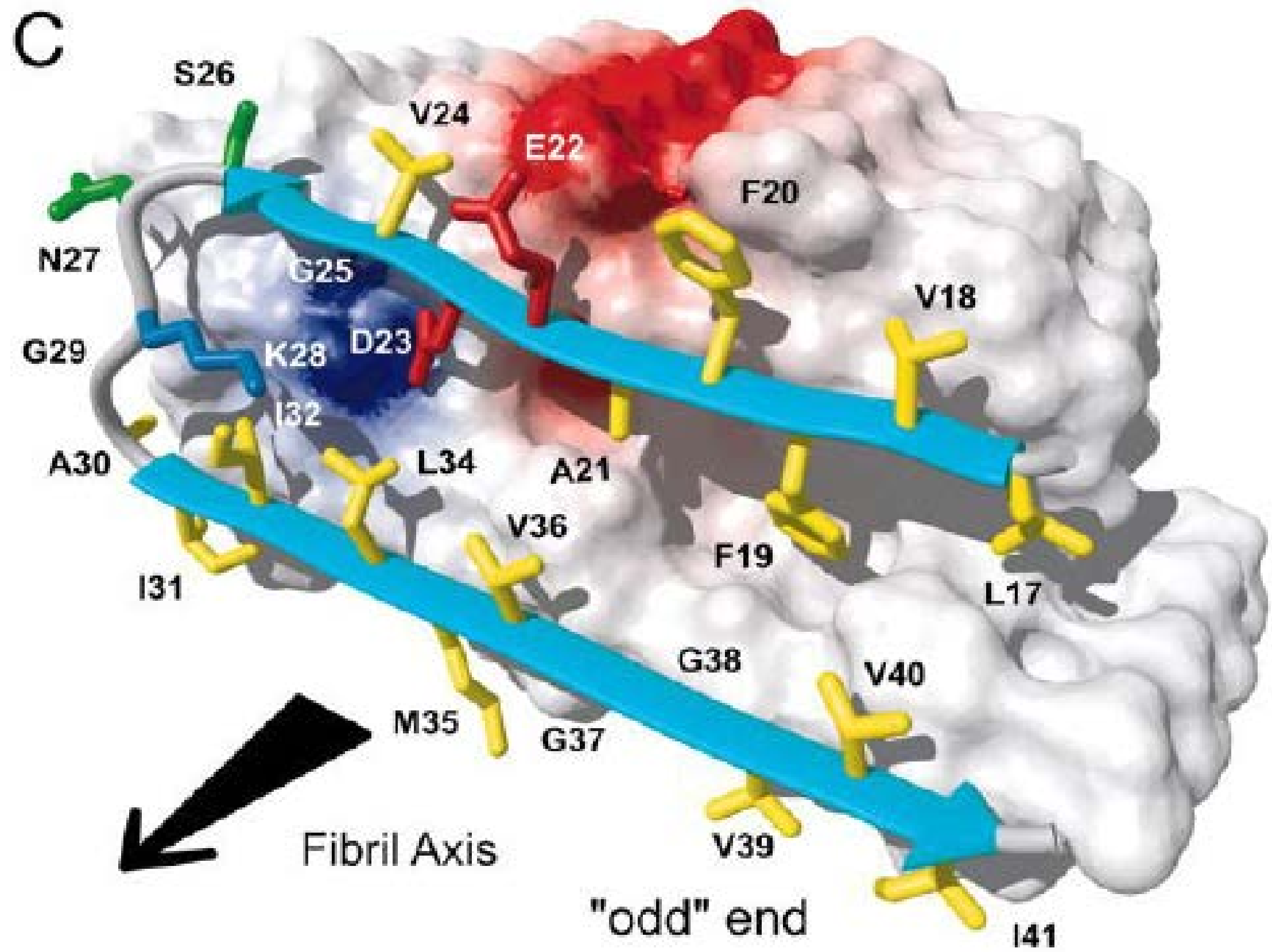

Fig. 1. Van der Waals contact surface polarity and ribbon diagram of the 17-42 residues of Aβ (from Fig. 4 of [1]). The 13-mer modular sandwiched β-sheets are indicated by cyan arrows, and the hydrophobic amino acid side chains are shown in yellow. Positively and negatively charged surface patches are shown in blue and red, respectively, with the D23-K28 (E22) salt bridge. The N terminal hydrophilic loop 1-16 (672-687), which connects successive fibrillar molecules, is disordered and is not resolved by NMR solution data.

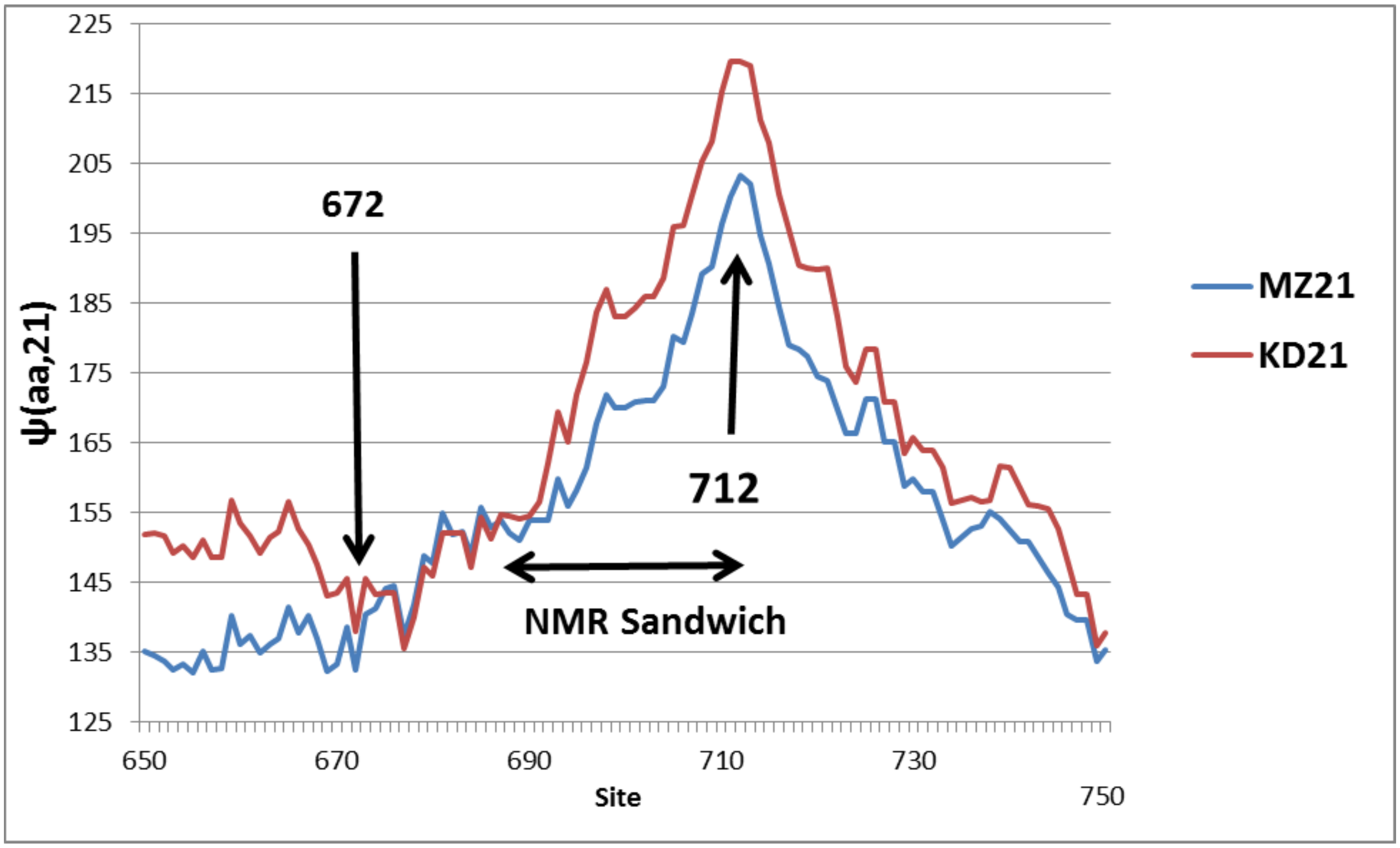

Fig. 2. Detail of the wide A4 hydrophobic peak of ψ(aa,21). The beginnings of the amyloid fragments occur just at the hydrophilic cleavage dip at Asp672, and end at the hydrophobic peak 712 ± 1. The flat top of KD21 spans 711-713. It may therefore be said that the KD scale correctly identifies the first-order separation of Aβ40 from Aβ42 (the splitting of 712 ± 1). It is possible that the 3 aa width of this KD flat peak is accurate because fragmentation occurs at a high effective temperature [17]. The range covered by the NMR sandwich model of Fig. 1 matches the abrupt increases of the MZ and KD ψ(aa,21) profiles near site 688.

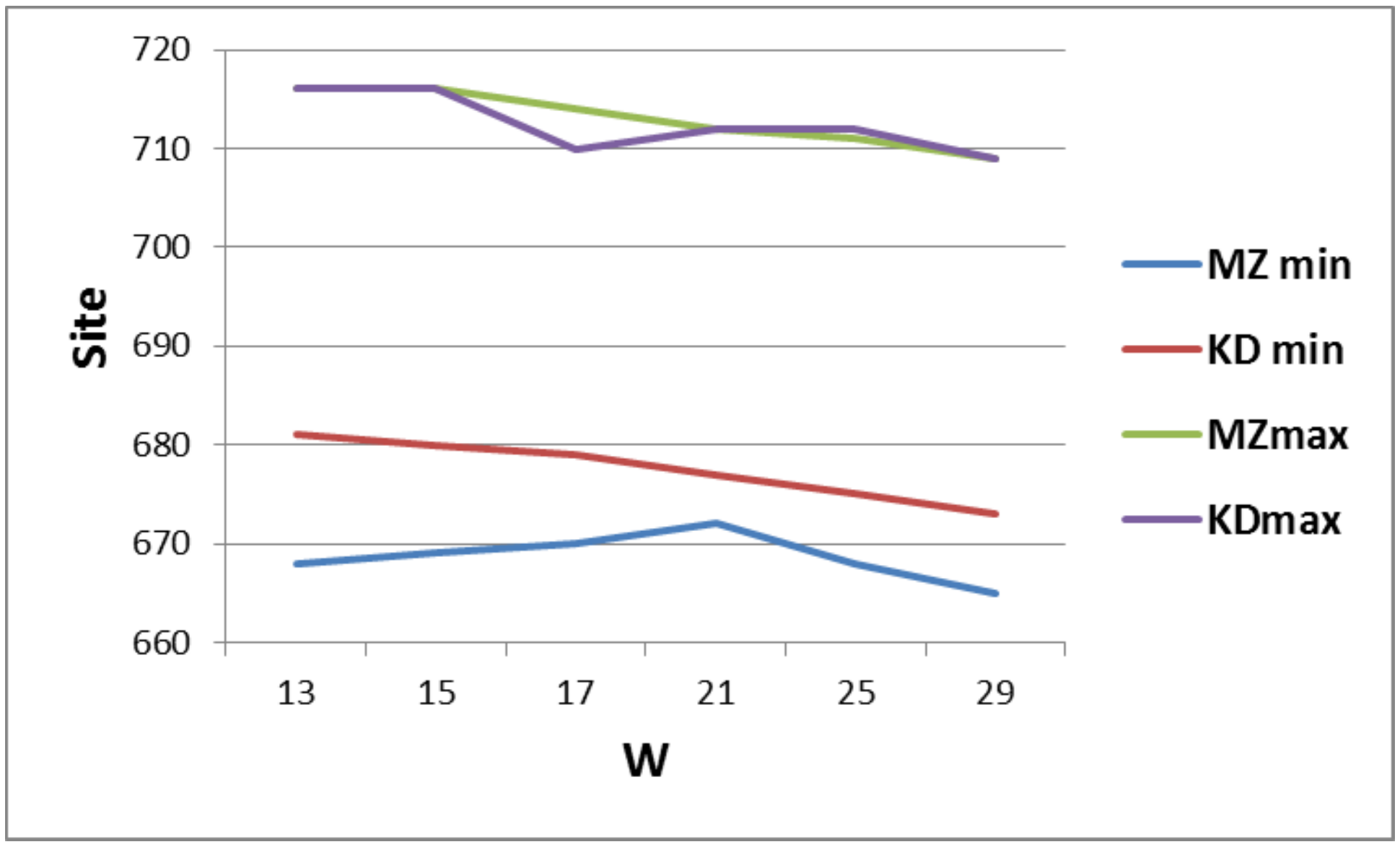

Fig. 3. Amphiphilic extrema of ψ(aa,W) profiles in the Aβ range. Fig. 2 showed the ψ(aa,21) profiles. This figure shows that the 672 “disordered” cleavage site is predicted correctly by hydrophilic maxima only with the MZ scale and W = 21. The 712 ± 1 cleavage site is predicted correctly by hydrophobic maxima with the KD scale for W = 21 and 25, and with the MZ scale for W = 21.

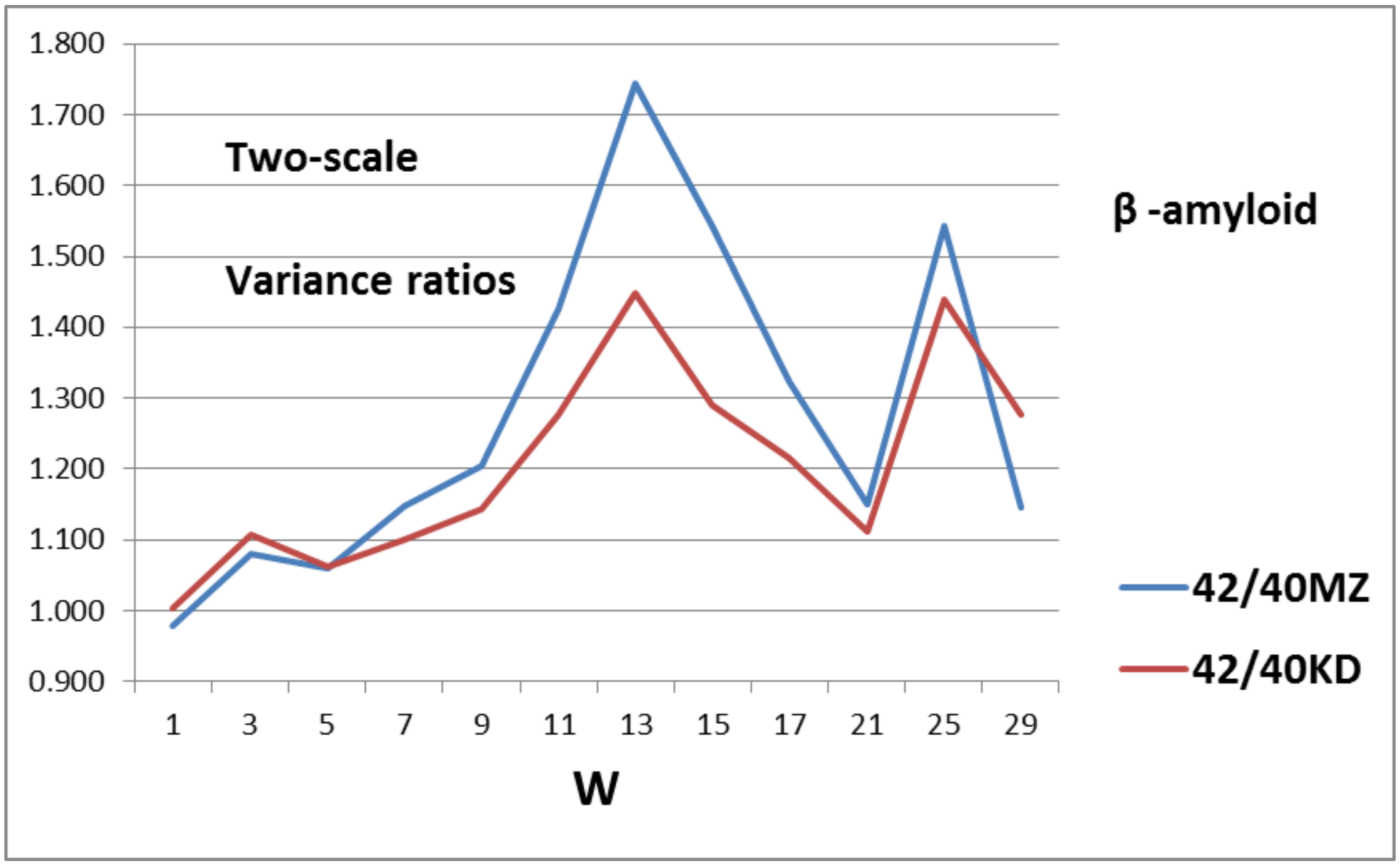

Fig. 4. The roughness or variance ℜ(W) = $<<\psi(aa,W)>^2> - <<\psi(aa,W)>>^2$ of amyloid β fragments as a function of the sliding window length W. Because Aβ 40 is much less toxic (aggregative) than Aβ 42, ℜ(W) of the latter is normalized by ℜ(W) of the former. Note that simply comparing the values of ℜ(1) fails, and that the W = 13 modularity peak of the sandwiched β-sheets is resolved well even with the KD scale. The MZ scale resolves thr modular structure better and produces a peak that is almost twice as large when W is tuned to 13.

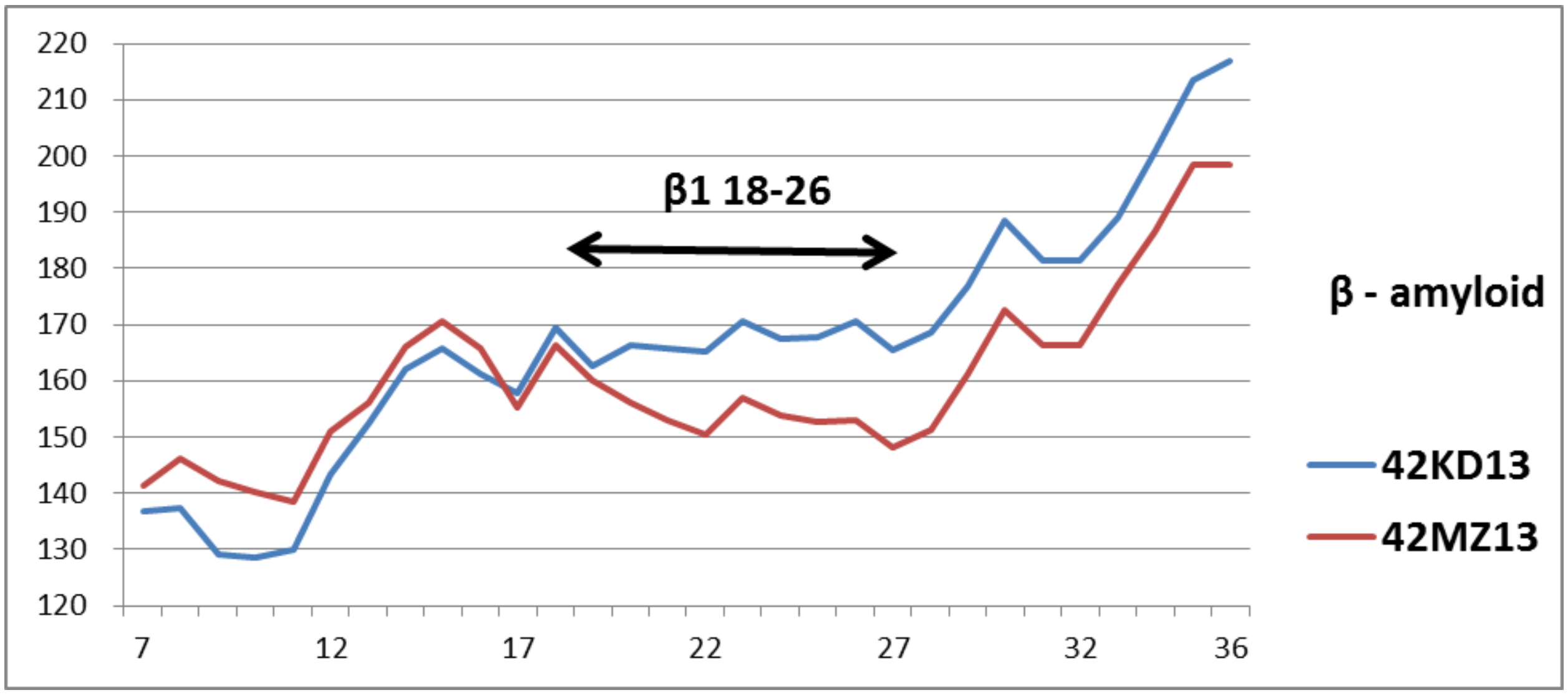

Fig. 5. Profiles of ψ(aa,13) for Aβ42 for the MZ SOC differential scale and the first-order KD scale, renumbered with 672 = 1. Note the hydrophilic softening of $\beta_1$ at the hairpin turn (26-30, see Fig. 1) with the SOC MZ scale. This $\beta_1$ softening facilitates formation of the modular sandwich structure, by making its central region more flexible. At the hairpin turn the W = 13 profile with the KD scale is nearly flat, but there is a local hydrophilicly soft minimum promoting a mechanical hinge in the MZ profile.